\documentclass[times,twocolumn,final]{elsarticle} 

\usepackage{medima}
\usepackage{framed,multirow}

\usepackage{amssymb}
\usepackage{latexsym}

\usepackage{url}
\usepackage{xcolor}
\usepackage{amsmath}
\usepackage{hyperref}
\usepackage{booktabs}
\usepackage{tabularx}
\usepackage{lipsum}
\usepackage{cleveref}
\usepackage{todonotes}
\usepackage{listings}

\newcommand{\blue}[1]{\color{blue}#1\color{black}}
\newcommand{\red}[1]{\color{red}#1\color{black}}
\definecolor{ao(english)}{rgb}{0.0, 0.5, 0.0}
\newcommand{\green}[1]{\color{ao(english)}#1\color{black}}
\usepackage[ruled,vlined]{algorithm2e}
\usepackage{color, colortbl} 
\usepackage[flushleft]{threeparttable} 
\usepackage{placeins} 
\usepackage{float}
\usepackage[normalem]{ulem}
\usepackage{enumitem}
\usepackage{changes} 
\setaddedmarkup{\green{#1}}
\setdeletedmarkup{ \red{\sout{#1}}}
\definecolor{newcolor}{rgb}{.8,.349,.1}

\journal{preprint}

\definecolor{C0}{rgb}{0.12156862745098039, 0.4666666666666667, 0.7058823529411765}
\definecolor{C1}{rgb}{1.0, 0.4980392156862745, 0.054901960784313725}
\definecolor{C2}{rgb}{0.17254901960784313, 0.6274509803921569, 0.17254901960784313}
\definecolor{C3}{rgb}{0.8392156862745098, 0.15294117647058825, 0.1568627450980392}
\definecolor{C4}{rgb}{0.5803921568627451, 0.403921568627451, 0.7411764705882353}
\definecolor{C5}{rgb}{0.5490196078431373, 0.33725490196078434, 0.29411764705882354}
\definecolor{C6}{rgb}{0.8901960784313725, 0.4666666666666667, 0.7607843137254902}
\definecolor{C7}{rgb}{0.4980392156862745, 0.4980392156862745, 0.4980392156862745}
\definecolor{C8}{rgb}{0.7372549019607844, 0.7411764705882353, 0.13333333333333333}
\definecolor{C9}{rgb}{0.09019607843137255, 0.7450980392156863, 0.8117647058823529}
 
\newcommand\panelnumstyle[1]{#1}
\newcommand\panelstyle[1]{\textit{#1}}

\newcommand\numbpMRIVisitsFullDatasetBeforeExclusion{9,275}
\newcommand\numbpMRIPatientsFullDatasetBeforeExclusion{7,430}
\newcommand\numbpMRIVisitsFullDataset{7,756}
\newcommand\numbpMRIPatientsFullDataset{6,380}
\newcommand\numbpMRIPatientsFullDatasetExcluded{1,519} 
\newcommand\numManualDelineationsFullDataset{3,050} 
\newcommand\numUnlabelledStudiesFullDataset{4,706} 
\newcommand\numManualDelineationsFirstSubset{100}
\newcommand\numUnlabelledStudiesFirstSubset{7,656}
\newcommand\numManualDelineationsSecondSubset{300}
\newcommand\numUnlabelledStudiesSecondSubset{7,456}
\newcommand\numManualDelineationsThirdSubset{1,000}

\newcommand\numManualDelineationsFirstSubsetPerSupervisedRun{80}
\newcommand\numManualDelineationsSecondSubsetPerSupervisedRun{240}
\newcommand\numManualDelineationsThirdSubsetPerSupervisedRun{800}
\newcommand\numManualDelineationsFullDatasetPerSupervisedRun{2,440}

\newcommand\numUnlabelledStudiesThirdSubset{6,756}

\newcommand\distPSALevelsRUMCShort{$8.0\ (5-11)$}
\newcommand\distAgeRUMCShort{$66\ (61-70)$}

\newcommand\manualAnnotationsStart{January 2016}
\newcommand\manualAnnotationsEnd{August 2018}
\newcommand\numManuallyAnnotatedLesions{1,315} 

\newcommand\numTrainingSamplesManualPIRADS{3,050} 

\newcommand\numZGTVisits{300}
\newcommand\numZGTPatients{300}
\newcommand\distPSALevelsZGTShort{$6.6\ (5-9)$} 
\newcommand\distAgeZGTShort{$65\ (59-68)$} 
\newcommand\ZGTDatasetStart{March 2015}
\newcommand\ZGTDatasetEnd{January 2017}
\newcommand\ZGTnumPatientsRP{61}
\newcommand\ZGTnumPatientsRPPercentage{$20.3\%$} 

\newcommand\numTrainingSamplesSanford{687}

\newcommand\numTrainingSamplesYuFPReduction{1,736} 

\newcommand\numTrainingSamplesBhattacharya{66} 

\newcommand\numTrainingSamplesNetzerBiop{806} 
\newcommand\numTrainingCasesHosseinzadehStart{50}
\newcommand\numTrainingCasesHosseinzadehEnd{1,586} 

\newcommand\numTrainingSamplesBiopMedian{146} 
\newcommand\numTrainingSamplesRadiolMedian{1,584} 
\newcommand\numIterationsPermutationTest{\replaced{$1,000,000$}{$100,000$}}
\newcommand\numIterationsBootstrapping{\replaced{$1,000,000$}{$100,000$}}

\newcommand\perfZGTAUCUNetagppManual{\replaced{$85.4\pm2.0\%$}{$84.6\pm2.0\%$}} 
\newcommand\perfZGTAUCUNetagppManualAndAutomatic{\replaced{$87.3\pm1.8\%$}{$86.3\pm1.8\%$}} 
\newcommand\perfZGTAUCUNetagppAutomaticSubsetOnly{\blue{$86.6\pm1.4\%$}} 
\newcommand\statZGTAUCUNetagppManualBetterThanManualAndAutomatic{$P=4.0\cdot 10^{-4}$} 
\newcommand\statZGTAUCUNetagppManualBetterThanAutomaticOnly{\blue{$P=4.4\cdot 10^{-4}$}} 
\newcommand\perfZGTAUCnnUNetManual{$87.2\pm0.8\%$} 
\newcommand\perfZGTAUCnnUNetManualAndAutomatic{$89.4\pm1.0\%$} 
\newcommand\perfZGTAUCnnUNetAutomaticSubsetOnly{\blue{$88.3 \pm 1.1\%$}} 
\newcommand\statZGTAUCnnUNetManualBetterThanManualAndAutomatic{$P < 10^{-4}$} 
\newcommand\statZGTAUCnnUNetManualBetterThanAutomaticOnly{\blue{$P=0.24$}} 

\newcommand\perfZGTSensitivityRadiologist{$93.2 \pm 2.7\%$} 
\newcommand\perfZGTSpecificityRadiologist{$75.9 \pm 2.9\%$} 
\newcommand\perfZGTSpecificityUnetagppManual{$35.8\pm11.3\%$} 
\newcommand\perfZGTSpecificityUnetagppManualAndAutomatic{$46.7 \pm 11.8\%$} 
\newcommand\statZGTSpecificityUNetagppManualBetterThanManualAndAutomatic{$P=7.8 \cdot 10^{-3}$} 
\newcommand\perfZGTSpecificitynnUNetManual{$51.5 \pm 8.1\%$} 
\newcommand\perfZGTSpecificitynnUNetManualAndAutomatic{$57.1 \pm 7.3\%$} 
\newcommand\statZGTSpecificitynnUNetManualBetterThanManualAndAutomatic{$P=0.28$} 

\newcommand\perfZGTpAUCUNetagppManual{$0.697 \pm 0.034$} 
\newcommand\perfZGTpAUCUNetagppManualAndAutomatic{$0.727 \pm 0.038$} 
\newcommand\perfZGTpAUCUNetagppAutomaticSubsetOnly{\blue{$2.024 \pm 0.039$}} 
\newcommand\statZGTpAUCUNetagppManualBetterThanManualAndAutomatic{$P=3.6\cdot 10^{-3}$} 
\newcommand\statZGTpAUCUNetagppManualBetterThanAutomaticOnly{\blue{$P < 10^{-4}$}} 
\newcommand\perfZGTpAUCnnUNetManual{$0.692 \pm 0.030$} 
\newcommand\perfZGTpAUCnnUNetManualAndAutomatic{$0.746 \pm 0.016$} 
\newcommand\perfZGTpAUCnnUNetAutomaticSubsetOnly{\blue{$1.949 \pm 0.056$}} 
\newcommand\statZGTpAUCnnUNetManualBetterThanManualAndAutomatic{$P<10^{-4}$} 
\newcommand\statZGTpAUCnnUNetManualBetterThanAutomaticOnly{\blue{$P=0.13$}} 

\newcommand\perfZGTSensAtOneFPUNetagppManual{$77.7 \pm 4.0\%$} 
\newcommand\perfZGTSensAtOneFPUNetagppManualAndAutomatic{$81.3 \pm 3.9\%$} 
\newcommand\perfZGTSensAtOneFPUNetagppAutomaticSubsetOnly{\blue{$83.5\pm1.9\%$}} 
\newcommand\statZGTSensAtOneFPUNetagppManualBetterThanManualAndAutomatic{$P=1.9 \cdot 10^{-3}$} 
\newcommand\statZGTSensAtOneFPUNetagppManualBetterThanAutomaticOnly{\blue{$P < 10^{-4}$}} 
\newcommand\perfZGTSensAtOneFPnnUNetManual{$76.4 \pm 3.8\%$} 
\newcommand\perfZGTSensAtOneFPnnUNetManualAndAutomatic{$83.6 \pm 2.3\%$} 
\newcommand\perfZGTSensAtOneFPnnUNetAutomaticSubsetOnly{\blue{$80.4\pm2.4\%$}}  
\newcommand\statZGTSensAtOneFPnnUNetManualBetterThanManualAndAutomatic{$P < 10^{-4}$} 
\newcommand\statZGTSensAtOneFPnnUNetManualBetterThanAutomaticOnly{\blue{$P=0.10$}}  

\newcommand\perfAutomaticFindingExtractionTotalCorrect{3,024} 
\newcommand\perfAutomaticFindingExtractionTotalIncorrect{20} 
\newcommand\perfAutomaticFindingExtractionTotalExcludedLabelled{8} 
\newcommand\perfAutomaticFindingExtractionTotalExcludedPercentage{$0.3\%$} 
\newcommand\perfAutomaticFindingExtractionTotalExcludedUnlabelled{121} 
\newcommand\perfAutomaticFindingExtractionTotalExcludedUnlabelledPercentage{$2.6\%$} 
\newcommand\perfAutomaticFindingExtractionTotal{3,044} 
\newcommand\perfAutomaticFindingExtractionAccuracy{$99.3\%$} 
\newcommand\perfAutomaticFindingExtractionErrorRate{$0.7\%$} 

\newcommand\perfLesionLocalisationnnUNetFPsUnfiltered{$0.064\pm 0.008$}
\newcommand\perfLesionLocalisationnnUNetFPsModelUnfiltered{$0.39\pm 0.14$}

\newcommand\perfLesionLocalisationUNetagppFPsUnfiltered{$0.097\pm 0.011$}
\newcommand\perfLesionLocalisationUNetagppFPsModelUnfiltered{$0.88\pm 0.29$}

\newcommand\numLesionsExcludednnUNet{119}
\newcommand\perfLesionLocalisationnnUNetSensitivityFiltered{$83.8\pm 1.1\%$}
\newcommand\perfLesionLocalisationnnUNetFPsFiltered{$0.063\pm 0.008$}

\newcommand\numLesionsExcludedUNetagpp{4}
\newcommand\perfLesionLocalisationUNetagppSensitivityFiltered{$78.5\pm 3.3\%$}
\newcommand\perfLesionLocalisationUNetagppFPsFiltered{$0.096\pm 0.012$}

\newcommand\perfLesionSegmentationAVAnnUNet{$0.51 \pm 0.33$} 
\newcommand\perfLesionSegmentationAVAnnUNetMatched{$0.67 \pm 0.19$} 

\newcommand\perfLesionSegmentationAVAUNetagpp{$0.45 \pm 0.28$} 
\newcommand\perfLesionSegmentationAVAUNetagppMatched{$0.57 \pm 0.17$} 

\newcommand\DSCPatientOneTop{0.70}

\newcommand\DSCPatientTwo{0.87}
\newcommand\DSCPatientThree{0.55}

\newcommand\RUMCAUCHosseinzadehStart{$79.9\%$}
\newcommand\RUMCAUCHosseinzadehEnd{$87.5\%$}

\newcommand\maxPvalueIterationOneBetterThanSupervised{$P < 10^{-4}$} 
\newcommand\PvalueIterationTwoBetterThanIterationOneFirstSubsetAUROC{$P = 3.9 \cdot 10^{-3}$} 
\newcommand\PvalueIterationTwoBetterThanIterationOneSecondSubsetAUROC{$P = 3.1 \cdot 10^{-3}$} 
\newcommand\numManualDelineationsMatchedPerformanceFullDatasetAUROC{169} 
\newcommand\numManualDelineationsMatchedPerformanceFullDatasetpAUC{431} 
\newcommand\annotationEfficiencyMatchedPerformanceFullDatasetAUROC{$14.4\times$} 
\newcommand\annotationEfficiencyMatchedPerformanceFullDatasetpAUC{$5.7\times$} 

\newcommand\PvalueSupervisedFullBetterThanThreeHundredIterationTwoAUROC{$P = 6.4 \cdot 10^{-3}$} 
\newcommand\PvalueOneHundredIterationTwoBetterThanSupervisedFullAUROC{$P = 0.014$} 
\newcommand\PvalueSupervisedFullBetterThanOneThousandIterationTwopAUC{$P < 10^{-4}$} 
\newcommand\PvalueThreeHundredIterationTwoBetterThanSupervisedFullpAUC{$P = 0.032$} 

\newcommand\nsig{$n_{\text{sig}}$}

\newcommand\nameUNet{\textit{U-Net}}
\newcommand\nameUNets{\textit{U-Nets}}
\newcommand\nameUNetagpp{\textit{DA-UNet}}
\newcommand\namennUNet{\textit{nnU-Net}}
\newcommand\nameLeakyReLU{\textit{LeakyReLU}}
\newcommand\nameRUMC{RUMC}
\newcommand\namepAUC{pAUC}
\newcommand\namepAUCs{pAUCs}

\begin{document}

\verso{Joeran Bosma \textit{et~al.}}

\begin{frontmatter}

\title{\replaced{Annotation-efficient cancer detection with report-guided lesion annotation \\for deep learning-based prostate cancer detection in bpMRI}{Report-guided automatic lesion annotation for deep learning-based prostate cancer detection in bpMRI}}

\author[1]{Joeran \snm{Bosma}\corref{cor1}}
\cortext[cor1]{Corresponding author: }
\ead{Joeran.Bosma@radboudumc.nl}
\author[2]{Anindo \snm{Saha}\fnref{1}}
\author[3]{Matin \snm{Hosseinzadeh}\fnref{1}}
\author[4]{Ilse \snm{Slootweg}\fnref{1}}
\author[5]{Maarten \snm{de Rooij}\fnref{1}}
\author[6]{Henkjan \snm{Huisman}\fnref{1}}

\address[1]{Diagnostic Image Analysis Group, Radboud University Medical Center, Nijmegen 6525 GA, The Netherlands}


\begin{abstract}
Deep learning-based diagnostic performance increases with more annotated data, \replaced{but large-scale manual annotations are expensive and labour-intensive}{but manual annotation is a bottleneck in most fields}. 
Experts evaluate diagnostic images during clinical routine, and write their findings in reports. 
\replaced{
Leveraging unlabelled exams paired with clinical reports \deleted{to train deep learning systems} could overcome the manual labelling bottleneck.
}{Automatic annotation based on clinical reports could overcome the manual labelling bottleneck.}
\replaced{We hypothesise that detection models can be trained semi-supervised with automatic annotations generated using model predictions, guided by sparse information from clinical reports.}{We hypothesise that dense annotations for detection tasks can be generated using model predictions, guided by sparse information from these reports. }
\replaced{To demonstrate efficacy, we train clinically significant prostate cancer (csPCa) segmentation models, where automatic annotations are}{To demonstrate efficacy, we generated clinically significant prostate cancer (csPCa) annotations,} guided by the number of clinically significant findings in the radiology reports. 
We included \numbpMRIVisitsFullDataset\ prostate MRI examinations, of which \numTrainingSamplesManualPIRADS\ were manually annotated\deleted{and \numUnlabelledStudiesFullDataset\ were automatically annotated}. 
\deleted{We evaluated the automatic annotation quality on the manually annotated subset: 
our score extraction correctly identified the number of csPCa lesions for \perfAutomaticFindingExtractionAccuracy\ of the reports and
our csPCa segmentation model correctly localised \perfLesionLocalisationnnUNetSensitivityFiltered\ of the lesions. }
We evaluated prostate cancer detection performance on \numZGTVisits\ exams from an external centre with histopathology-confirmed ground truth. 
\replaced{Semi-supervised training}{Augmenting the training set with automatically labelled exams} improved case-based diagnostic area under the receiver operating characteristic curve \added{(AUROC)} from \perfZGTAUCnnUNetManual\ to \perfZGTAUCnnUNetManualAndAutomatic\ (\statZGTAUCnnUNetManualBetterThanManualAndAutomatic) 
and improved lesion-based sensitivity at one false positive per case from \perfZGTSensAtOneFPnnUNetManual\ to \perfZGTSensAtOneFPnnUNetManualAndAutomatic\ (\statZGTSensAtOneFPnnUNetManualBetterThanManualAndAutomatic)\deleted{, with $mean \pm std.$ over 15 independent runs}. 
Semi-supervised training was \annotationEfficiencyMatchedPerformanceFullDatasetAUROC\ more annotation-efficient for case-based performance and \annotationEfficiencyMatchedPerformanceFullDatasetpAUC\ more annotation-efficient for lesion-based performance.
%
This improved performance demonstrates the feasibility of our \replaced{training procedure}{report-guided automatic annotations}. 
Source code is \deleted{made} publicly available at \href{https://github.com/DIAGNijmegen/Report-Guided-Annotation}{github.com/DIAGNijmegen/Report-Guided-Annotation}. Best csPCa detection algorithm is \deleted{made} available at \href{https://grand-challenge.org/algorithms/bpmri-cspca-detection-report-guided-annotations/}{grand-challenge.org/algorithms/bpmri-cspca-detection-report-guided-annotations/}. 


\end{abstract}


\end{frontmatter}


\section{Introduction}
Computer-aided diagnosis (CAD) systems in fields where clinical experts are matched or outperformed, typically use very large training datasets. 
Top performing deep learning systems used 29,541 training cases (10,306 patients) for the detection of lung cancer \citep{ardila2019end}, 121,850 training cases (121,850 women) for the detection of breast cancer \citep{mckinney2020international} and 16,114 training cases (12,399 patients) for the classification of skin diseases \citep{liu2020deep}. 

Annotation time and cost are major limiting factors, resulting in significantly smaller labelled training datasets in most deep learning fields. 
In the natural image domain, leveraging samples without target task annotations has proven to be effective, even when manually labelled samples are abundant. 
On ImageNet, with 1.3 million manually labelled training samples, all \replaced{ten}{eight} leaderboard holders of the past \replaced{four}{three} years used additional training samples with automatically generated labels. 
Several approaches were used: \citep{mahajan2018exploring} used 3.5 billion images from Instagram to pre-train their models by predicting the corresponding hashtag (\textit{transfer learning}); 
\citep{xie2020self} used a \textit{teacher} model to predict 300 million images scraped from the web and selected 130 million samples to train a new \textit{student} model (\textit{self-training}); 
and \citep{pham2020meta} pushed the teacher-student approach further by continuously updating the teacher model with reinforcement learning. 
Although the other \replaced{seven}{five} leaderboard holders primarily pursued different research directions, they did incorporate automatic labelling techniques to reach state-of-the-art performance.

In the medical domain, popular techniques to leverage unlabelled samples include transfer learning from a distant or related task, and self-training with automatically generated labels \citep{cheplygina2019not}. 
Other techniques to leverage unlabelled samples include contrastive learning \citep{chaitanya2020contrastive, sowrirajan2021moco, azizi2021big} and self-supervised representation learning \citep{zhou2019models}. These techniques either pre-train without labels, or directly use model predictions as true labels. 

Leveraging clinical information, which is often available in medical reports, to improve training with unlabelled samples is under-explored. 
Clinical information from reports typically differ from regular training annotations, but can inform the generation of automatic annotations for self-training. 
One study, \citep{bulten2019automated}, generated pixel-level Gleason score annotations in H\&E stained prostate biopsies by leveraging pathology reports. First, they generated precise cancer masks. Then, they extracted the Gleason scores from the pathology reports to classify the cancer masks into Gleason grades. These steps allowed the generation of Gleason score annotations in thousands of prostate biopsies, which would have been infeasible to obtain manually. 
Incorporating clinical information to guide automatic annotations for self-training remains to be investigated for medical tasks other than biopsy grading. 

We hypothesise that medical detection tasks, where the structure\added{s} of interest can be counted, can leverage unlabelled cases \replaced{by}{using} \added{semi-supervised training with} report-guided \deleted{automatic} annotations. 
Specifically, we focus on lesion detection, where each case can have any number of lesions. 
To demonstrate feasibility of our method, we developed \replaced{a semi-supervised training}{an automatic annotation} procedure for clinically significant prostate cancer detection in MRI. 

Prostate cancer (PCa) has 1.2 million new cases each year \citep{GCS2020}, a high incidence-to-mortality ratio and risks associated with treatment and biopsy; making non-invasive diagnosis of clinically significant prostate cancer (csPCa) crucial to reduce both overtreatment and unnecessary (confirmatory) biopsies \citep{stavrinides2019mri}. Multiparametric MRI (mpMRI) scans interpreted by expert prostate radiologists provide the best non-invasive diagnosis \citep{eldred2021population}, but is a limited resource that cannot be leveraged freely. 
%
Computer-aided diagnosis (CAD) can assist radiologists to diagnose csPCa, but present-day solutions lack stand-alone performance comparable to that of expert radiologists \citep{cao2021performance, saha2021end, hosseinzadeh21deep, schelb2019classification, seetharaman2021automated}. 

Datasets used for prostate cancer detection and diagnosis have significantly fewer training samples than datasets used to train top-performing deep learning systems in other medical fields \citep{ardila2019end, mckinney2020international, liu2020deep}. 
Studies tackling csPCa detection in MRI by training on histopathology-confirmed annotations, used datasets with \numTrainingSamplesBhattacharya -\numTrainingSamplesNetzerBiop\ (median: \numTrainingSamplesBiopMedian) samples to train their deep learning system \citep{schelb2019classification, arif2020clinically, sanyal2020automated, aldoj2020semi, bhattacharya2020corrsignet, cao2019joint, seetharaman2021automated, netzer2021fully}. 
Approaches using radiologically-estimated annotations (reported using Prostate Imaging Reporting and Data System: Version 2 (PI-RADS)) used \numTrainingSamplesSanford -\numTrainingSamplesYuFPReduction\ (median: \numTrainingSamplesRadiolMedian) training samples \citep{sanford2020deep, yu2020deep, yu2020false, saha2021end, hosseinzadeh21deep}. 


Prior work investigated the effect of training set size on prostate cancer detection performance, with radiologically-estimated ground truth for training and testing \citep{hosseinzadeh21deep}. This work shows patient-based area under the receiver operating characteristic curve (AUROC) for their internal test set increased logarithmically between \numTrainingCasesHosseinzadehStart\ and \numTrainingCasesHosseinzadehEnd\ training cases, from an AUROC of \RUMCAUCHosseinzadehStart\ to \RUMCAUCHosseinzadehEnd. 
If this trend continues, tens of thousands of annotated cases would be required to reach expert performance — in concordance with similar applications in medical imaging. 

Trained investigators supervised by an experienced radiologist annotated all PI-RADS $\geq 4$ findings in more than three thousand of our institutional prostate MRI exams. According to our principal annotator, I.S., 
she requires about four minutes to annotate a single prostate cancer lesion in 3D. 
Difficult cases are discussed with radiologists, further increasing the overall duration. Annotating tens of thousands of cases would therefore incur huge costs and an incredibly large time investment. 


\replaced{Our report-guided semi-supervised training procedure aims to leverage unlabelled exams paired with clinical reports, to improve detection performance without any additional manual effort. 
We investigate the efficacy of our training procedure by comparing semi-supervised training of csPCa detection models against supervised training. 
First, we investigate the setting with \numManualDelineationsFullDataset\ manually annotated exams and \numUnlabelledStudiesFullDataset\ unlabelled exams. 
Secondly, we investigate several labelling budgets: \numManualDelineationsFirstSubset, \numManualDelineationsSecondSubset, \numManualDelineationsThirdSubset\ or \numManualDelineationsFullDataset\ manually annotated exams, paired with the remaining \numUnlabelledStudiesFirstSubset, \numUnlabelledStudiesSecondSubset, \numUnlabelledStudiesThirdSubset\ or \numUnlabelledStudiesFullDataset\ unlabelled exams, respectively. 
Finally, to determine annotation-efficiency, we compare supervised training performance with \numManualDelineationsFullDataset\ manually annotated exams against semi-supervised training with reduced manual annotation budgets. 
}{Our automatic labelling procedure aims to leverage unlabelled cases without any additional manual effort, to use the largest dataset of MRI scans for prostate cancer detection reported in literature to date. 
We investigate the efficacy of our automatically generated annotations by training csPCa detection models on manually and/or automatically annotated exams. 
We compare performance of these models to investigate how automatic annotations compare to manual annotations, and what automatically annotated exams add to manually annotated exams. }

\replaced{In this study, the training procedure with report-guided automatic annotations}{The report-guided automatic annotation procedure}
is presented for csPCa detection in bpMRI using radiology reports. However, the underlying method is neither limited to csPCa, MRI, nor radiology reports, and can be applied universally. 
Any detection task with countable structures of interest, and clinical information reflecting these findings, 
\replaced{can use our training method to leverage unlabelled exams with clinical reports}{can leverage our method to automatically annotate examinations}. 
\section{Materials and Methods}
\label{sec:methods}

\subsection{Datasets}
Two datasets with biparametric MRI (bpMRI) scans (axial T2-weighted (T2W), high b-value ($\geq 1400$) diffusion-weighted imaging (DWI) and apparent diffusion coefficient (ADC) maps) for prostate cancer detection were used. 

To train and tune our models, \numbpMRIVisitsFullDataset\ studies (\numbpMRIPatientsFullDataset\ patients) out of \numbpMRIVisitsFullDatasetBeforeExclusion\ consecutive studies (\numbpMRIPatientsFullDatasetBeforeExclusion\ patients) from \deleted{the} Radboud University Medical Centre (\nameRUMC) were included. \numbpMRIPatientsFullDatasetExcluded\ studies were excluded due to 
incomplete examinations, 
preprocessing errors, 
prior treatment, 
poor scan quality, 
or a prior positive biopsy (Gleason grade group $\geq 2$). 
All scans were obtained as part of clinical routine and evaluated by at least one of six experienced radiologists (4–25 years of experience\added{ with prostate MRI}). 
All \numManuallyAnnotatedLesions\ csPCa lesions (PI-RADS $\geq4$) in \numManualDelineationsFullDataset\ studies between \manualAnnotationsStart\ and \manualAnnotationsEnd\ were manually delineated by trained investigators (at least 1 year of experience), supervised by an experienced radiologist (M.R., 7 years of experience with prostate MRI). 



To test our models, an external dataset of \numZGTVisits\ exams (\numZGTVisits\ patients) from Ziekenhuisgroep Twente (ZGT)\replaced{ was used, acquired between \ZGTDatasetStart\ and \ZGTDatasetEnd}{, acquired between \ZGTDatasetStart\ and \ZGTDatasetEnd\ was used}. All patients in the test set received TRUS biopsies and patients with suspicious findings on MR also received MR-guided biopsies. 
\replaced{%
For \ZGTnumPatientsRP\ exams (\ZGTnumPatientsRPPercentage) the ground truth was derived from radical prostatectomy, which superseded the biopsy findings. All examinations in the test set have histopathology-confirmed ground truth, while retaining the patient cohort observed in clinical practice. 
}{%
For \ZGTnumPatientsRP/\numZGTVisits\ (\ZGTnumPatientsRPPercentage) exams the ground truth was derived from radical prostatectomy, resulting in histopathology-confirmed ground truth for all examinations in the test set.
}



Further details on patient demographics, study inclusion/exclusion criteria and acquisition parameters can be found in the Supplementary Materials.
\subsection{\replaced{Report-guided Automatic Annotation}{\\Automated Segmentation of Report Findings}}


Radiology reports were used to automatically create voxel-level annotations for csPCa. At a high level, our labelling procedure consists of two steps:
\begin{enumerate}
    \item Count the number of clinically significant findings in each radiology report,
    \item Localise these findings in their corresponding bpMRI scans with a prostate cancer segmentation model.
\end{enumerate}

\begin{figure*}[t!]
    \centering
    \includegraphics[width=.775\textwidth]{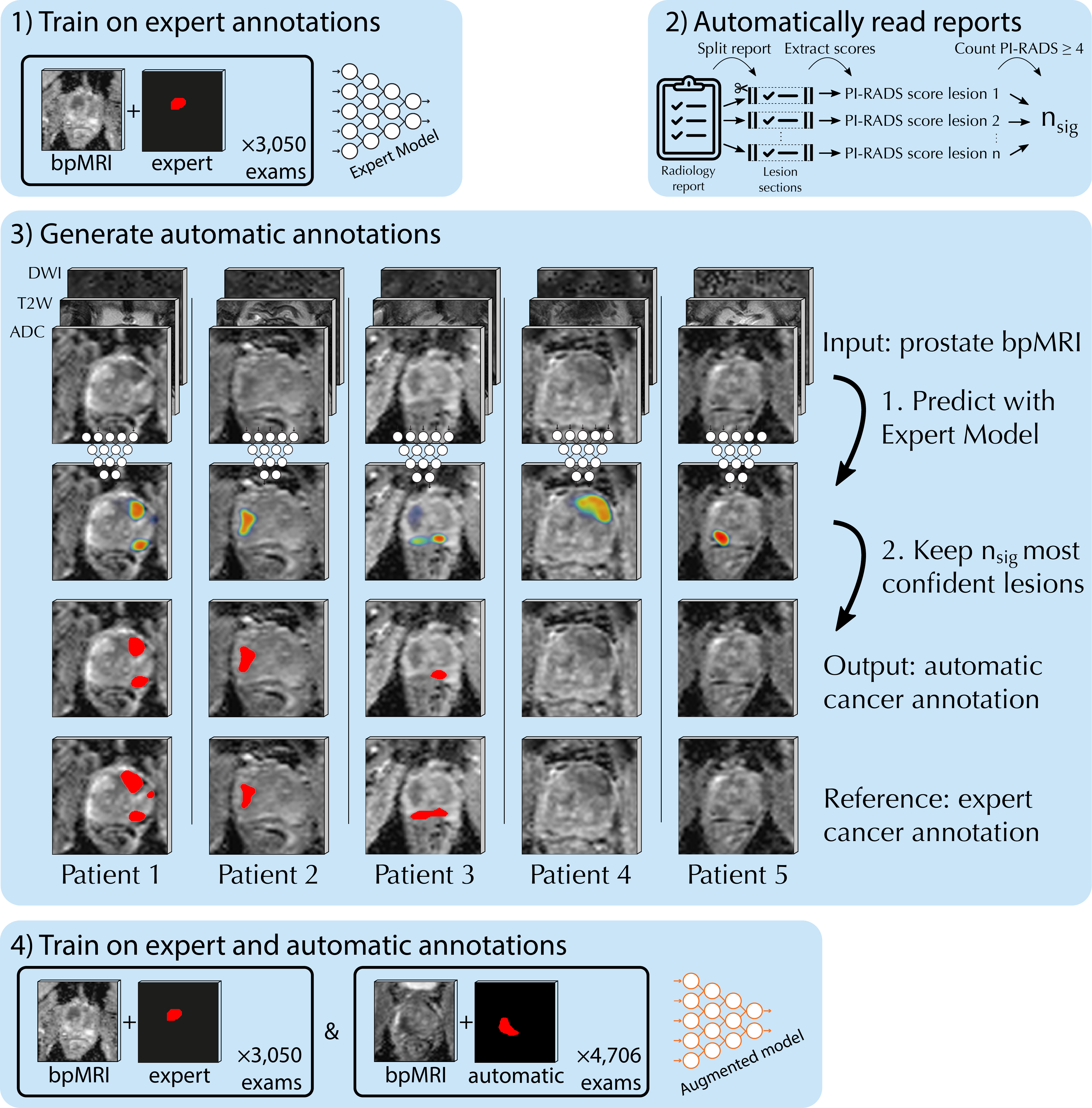}
    \caption{Overview of the steps to create automatic annotations for the unlabelled \nameRUMC\ exams. 
    \panelnumstyle{1)}\ Train a prostate cancer segmentation model on manually annotated clinically significant prostate cancer lesions (csPCa, PI-RADS $\geq 4$), the Expert model. 
    \panelnumstyle{2)}\ Extract the PI-RADS scores from the radiology reports and count the number of csPCa lesions, \nsig. 
    \panelnumstyle{3)}\ Localise and segment the csPCa lesions, by keeping the \nsig\ most confident lesion candidates of the Expert Model. 
    \panelnumstyle{4)}\ Automatic annotations are used to augment the training dataset and train a new prostate cancer segmentation model. 
    }
    \label{fig:pseudo-label-pipeline}
\end{figure*}

A rule-based natural language processing script was developed to automatically extract the PI-RADS scores from radiology reports.
The number of clinically significant findings, \nsig, is then defined as the number of PI-RADS $\geq 4$ findings in an exam. 
The clinically significant findings are localised by keeping the \nsig\ most confident candidates from a csPCa segmentation model, as depicted in \Cref{fig:pseudo-label-pipeline}. 
These automatically generated voxel-level masks can be used to augment the training dataset and produce new csPCa segmentation models. 


\subsubsection{Extraction of Report Findings}
\label{sec:automatic_finding_extraction}
%
%
Most of the radiology reports in our dataset were generated from a template, and modified to provide additional information. Although multiple templates were used over the years, 
this resulted in structured reports for most exams. 
This makes a rule-based natural language processing script a reliable and transparent way to extract PI-RADS scores from our radiology reports. 

Simply counting the occurrences of `PI-RADS 4/5' in the report body is reasonably effective, but has some pitfalls. 
For example, prior PI-RADS scores are often referenced during follow-up exams, resulting in false positive matches. 
Findings can also be grouped and described jointly, resulting in false negatives. 
To improve the reliability of the PI-RADS extraction from radiology reports, we extracted the scores in two steps. 


First, we tried to split the radiology reports in sections for individual findings. Secondly, we extracted the PI-RADS scores for each section individually. In case the report could not be split in sections per lesion, we applied strict pattern matching on the full report. See the Supplementary Materials for more details. 
%
Example report sections 
are shown in \Cref{fig:finding_extraction_radiology_report}. 

\begin{figure*}[ht!]
\centering
\noindent\fcolorbox{gray!8}{gray!8}{%
\begin{tabular}{c|c}
    \small\fontfamily{raleway}\selectfont \begin{minipage}[c]{0.47\textwidth}\input{figures/report_text/sample-1}\vspace{\belowcaptionskip}\end{minipage} & %
    \small\fontfamily{raleway}\selectfont \begin{minipage}[c]{0.47\textwidth}\input{figures/report_text/sample-2}\vspace{\belowcaptionskip}\end{minipage} \\
\end{tabular}
}
\caption{Example lesion report sections. The rule-based score extraction matched the \color{C1}T2W\color{black}, \color{C2}DWI \color{black}and \color{C3}DCE \color{black}scores coloured \color{C1}orange\color{black}, \color{C2}green \color{black}and \color{C3}red\color{black}, respectively. The resulting \color{C4}PI-RADS \color{black}score is coloured \color{C4}purple\color{black}. The reports were split in sections by matching the \color{C0}lesion identifier \color{black}in \color{C0}blue\color{black}. All reports were originally Dutch. }
\label{fig:finding_extraction_radiology_report}
\end{figure*}

\subsubsection{Localisation of Report Findings}
\label{sec:methods_localisation_report_findings}
\begin{figure}[h!]
    \centering
    \includegraphics[width=\columnwidth]{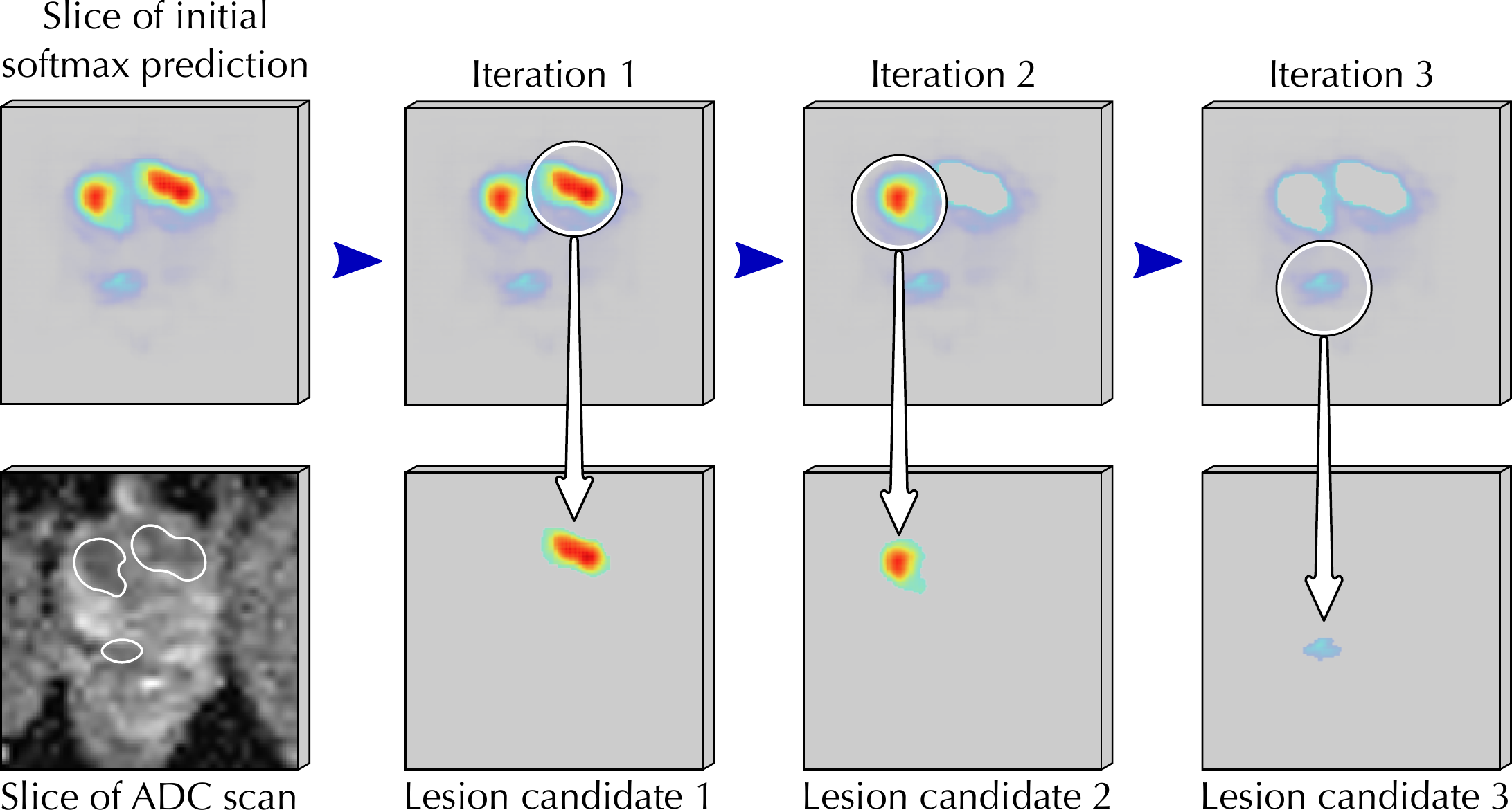}
    \caption{Depiction of the dynamic lesion candidate selection from a voxel-level prediction. The selected slice shows two high-confidence and one low-confidence lesion candidate extracted from the initial voxel-level prediction. All steps are performed in 3D. }
    \label{fig:dynamic-threshold-pipeline}
\end{figure}

To localise csPCa findings in unlabelled bpMRI scans, we employed an ensemble\footnote{Multiple models were ensembled by averaging the softmax confidence maps, which resulted in more consistent segmentation masks compared to a single model. The ensemble also improved localisation of report findings in difficult cases, where a single model was more likely to miss the lesion. } of csPCa segmentation models\deleted{ trained on manually annotated bpMRI scans}. From the resulting voxel-level confidence maps we created distinct lesion candidates, as illustrated in \Cref{fig:dynamic-threshold-pipeline}. Specifically, we create\added{d} a lesion candidate by starting at the most confident voxel, and including all connected voxels (in 3D) with at least $40\%$ of the peak's confidence. 
\replaced{Then, the candidate lesion is removed from the confidence map, and the process is repeated until no candidates remain or a maximum of 5 lesions are extracted.}{After a lesion candidate is extracted, we remove it from the model prediction and continue to the next peak.} \added{Tiny candidates of 10 or fewer voxels ($\leq 0.009\ \text{cm}^3$) are discarded.}

Automatic voxel-level csPCa annotations were generated by keeping the \nsig\ most confident lesion candidates, with \nsig\ the number of clinically significant report findings as described in \Cref{sec:automatic_finding_extraction}. 
If there were fewer lesion candidates than clinically significant report findings, the automatic label is excluded. 


\subsection{Models, Preprocessing and Data Augmentation}
\label{sec:model}

We have posed the prostate cancer detection task as a voxel-level segmentation task and employed two independent architectures, \namennUNet\ and Dual-Attention U-Net (\nameUNetagpp). 
\namennUNet\ is a self-configuring framework that follows a set of rules to select the appropriate architecture based on the input dataset \citep{isensee2021nnu}.
The \nameUNetagpp\ architecture was developed specifically for csPCa detection, where it performed best among state-of-the-art architectures \citep{saha2021end}. 
Both are derived from the \nameUNet\ architecture \citep{ronneberger2015u}, extended to 3D \citep{cciccek20163d}, and use anisotropic pooling or convolutional strides to account for the difference in through-plane and in-plane resolution of prostate MRI. 
See the Supplementary Materials for more details. 

The \namennUNet\ framework typically uses the sum of cross-entropy and soft Dice loss, and applies the loss at multiple resolutions (deep supervision). Motivated by \cite{baumgartner2021nndetection} and exploratory experiments, we trained \namennUNet\ using cross-entropy only. 
Based on prior experience \citep{saha2021end}, we trained the \nameUNetagpp\ model with Focal Loss ($\alpha=0.75$) \citep{lin2017focal}. 

The acquisition protocol of bpMRI ensures negligible movement between imaging sequences, and little deviation of the prostate from the centre of the scan. Therefore, neither registration between sequences, nor centring of the prostate was deemed necessary. 
In previous work, we have observed that a centre crop size of $72.0 \text{ mm}\times 72.0 \text{ mm}\times 64.8 \text{ mm}$ at a resampled resolution of $0.5\text{ mm}\times 0.5\text{ mm}\times 3.6\text{ mm/voxel}$ works well for our dataset \citep{saha2021end}. 
To prevent the \namennUNet\ framework to zero-pad these scans, we extended the field of view slightly for this model, to $80.0\text{ mm}\times  80.0\text{ mm}\times 72.0\text{ mm}$\deleted{ at a resampled resolution of $0.5\text{ mm}\times 0.5\text{ mm}\times 3.6\text{ mm/voxel}$}, which corresponds to a matrix size of $160\times160\times20$. 

\namennUNet\ comes with a predefined data preprocessing and augmentation pipeline, detailed in Supplementary Notes 2.2, 3.2 and 4 of \citep{isensee2021nnu}. 
In short, T2W and DWI scans undergo instance-wise \textit{z}-score normalisation, while ADC maps undergo robust, global \textit{z}-score normalisation with respect to the complete training dataset. 
For our anisotropic dataset, the \namennUNet\ framework applies affine transformations in 2D and applies a wide range of intensity and structure augmentations (Gaussian noise, Gaussian blur, brightness, contrast, simulation of low resolution and gamma augmentation). 


For the \nameUNetagpp\ model we used our institutional augmentation pipeline. 
We perform instance-wise min/max normalisation for T2W and DWI scans. For ADC maps, we divide each scan by 3000 ($97.4^{\text{th}}$ percentile), to retain their diagnostically relevant absolute values. 
Then, we apply Rician noise \citep{gudbjartsson1995rician} with $\sigma = 0.01$ to each scan at original resolution, with a probability of $75\%$. 
Subsequently, we resample all scans to a uniform resolution of 
$0.5\text{ mm}\times 0.5\text{ mm}\times 3.6\text{ mm/voxel}$ 
with bicubic interpolation. 
Finally, with a probability of $50\%$, we apply 2D affine data augmentations: horizontal mirroring, rotation with $\theta \sim 7.5\cdot\mathcal{N}(0, 1)$, horizontal translation with $h_x \sim 0.05\cdot\mathcal{N}(0, 1)$, vertical translation with $h_y \sim 0.05\cdot\mathcal{N}(0, 1)$ and zoom with $s_{xy} \sim 1.05\cdot\mathcal{N}(0, 1)$, where $\mathcal{N}(0, 1)$ is a Gaussian distribution with zero mean and unit variance. 
\subsection{Experimental Analysis}
\subsubsection{Extraction of Report Findings}
Accuracy of automatically counting the number of PI-RADS $\geq 4$ lesions in a report (\nsig) is determined by comparing against the number of PI-RADS $\geq 4$ lesions in the manually annotated \nameRUMC\ dataset. 
To account for multifocal lesions (which can be annotated as two distinct regions or a single larger one) and human error in the ground truth annotations, we manually checked the radiology report and verified the number of lesions when there was a mismatch between the ground truth and automatic estimation. 

\subsubsection{Localisation of Report Findings}
Localisation of clinically significant report findings 
is evaluated with the sensitivity and average number of false positives per case. Evaluation is performed with 5-fold cross-validation on the labelled \nameRUMC\ dataset, 
for which PI-RADS $\geq 4$ lesions were manually annotated in the MRI scan. 

\subsubsection{Segmentation of Report Findings}
Quality of the correctly localised report findings is evaluated with the Dice similarity coefficient (DSC). This evaluation is performed with 5-fold cross-validation on the labelled \nameRUMC\ dataset, 
which has manual PI-RADS $\geq 4$ lesion annotations. 

\subsubsection{Prostate Cancer Detection and Statistical Test}
\label{sec:model_evaluation}
Prostate cancer detection models 
are evaluated on \numZGTVisits\ external exams from ZGT, with histopathology-confirmed ground truth for all patients. 
Studies are considered positive if they have at least one Gleason grade group $\geq 2$ lesion (csPCa) \citep{epstein20162014}. 
Patient-based diagnostic performance was evaluated using the Receiver Operating Characteristic (ROC), and summarised to the area under the ROC curve (AUROC). 
Lesion-based diagnostic performance was evaluated using Free-Response Receiver Operating Characteristic (FROC), and summarised to the partial area under the FROC curve (\namepAUC) between \replaced{0 and 1}{0.01 and 2.50} false positive\deleted{s} per case, similar to \cite{saha2021end}. 
We trained our models with 5-fold cross-validation and 3 restarts for \namennUNet\ and 5 restarts for \nameUNetagpp, resulting in 15 or 25 independent AUROCs and \namepAUCs\ on the test set for each model configuration. 
To determine the probability of one configuration outperforming another configuration, we performed a permutation test with \numIterationsPermutationTest\ iterations. 
We used a statistical significance threshold of $0.01$.

$95\%$ confidence intervals (CI) for the radiologists were determined by bootstrapping \numIterationsBootstrapping\ iterations, with each iteration selecting $\sim\mathcal{U}(0, N)$ samples with replacement and calculating the target metric. Iterations that sampled only one class were rejected. 


\subsubsection{\added{Annotation-efficiency}}
\added{Annotation-efficiency of semi-supervised training is defined as the fraction of manual annotations that are required to reach the same performance as with supervised training. With $N_{supervised}$ the number of manually annotated exams used for supervised training and $N_{semi-supervised}$ the number of manually annotated exams required to reach the same performance, we obtain the annotation-efficiency ratio:
\begin{equation}
R = \frac{N_{supervised}}{N_{semi-supervised}}
\end{equation}

To construct a continuous curve for semi-supervised performance as function of the number of manually annotated exams, the performance of manual annotation budgets is piecewise logarithmically interpolated, as illustrated by the coloured dashed lines in the bottom row of \Cref{fig:ZGT_four} (which appear linear with a logarithmic x-axis). The required number of manual annotations is then derived from the intersections, as illustrated by the black dashed lines in the bottom row of \Cref{fig:ZGT_four}. This evaluates to:
\begin{equation}
    N_{semi-supervised} = N_a \cdot \left(\frac{N_b}{N_a}\right)^{(perf_{supervised}-perf_a)/(perf_b-perf_a)}
\end{equation}
Where $N_a$ is the number of manually annotated exams for the budget with performance just below supervised training, $N_b$ is the number of manually annotated exams for the budget with performance just above supervised training, $perf_{supervised}$ is the performance from supervised training, $perf_a$ is the performance with semi-supervised training just below supervised training and $perf_b$ is the performance with semi-supervised training just above supervised training. 

Manual annotation budgets of \numManualDelineationsFirstSubset, \numManualDelineationsSecondSubset, \numManualDelineationsThirdSubset\ or \numManualDelineationsFullDataset\ manually annotated exams, paired with the remaining \numUnlabelledStudiesFirstSubset, \numUnlabelledStudiesSecondSubset, \numUnlabelledStudiesThirdSubset\ or \numUnlabelledStudiesFullDataset\ unlabelled exams were investigated. 
For supervised training with 5-fold cross validation, this corresponds to \numManualDelineationsFirstSubsetPerSupervisedRun, \numManualDelineationsSecondSubsetPerSupervisedRun, \numManualDelineationsThirdSubsetPerSupervisedRun\ or \numManualDelineationsFullDatasetPerSupervisedRun\ manually annotated training samples per run. The supervised models are ensembled to generate automatic annotations for semi-supervised training, so for semi-supervised training the full manual annotation budgets are required. 
}
\section{Results}
\label{sec:results}

\subsection{Extraction of Report Findings}
\begin{figure}[h!]
    \centering
    \includegraphics[width=.9\columnwidth]{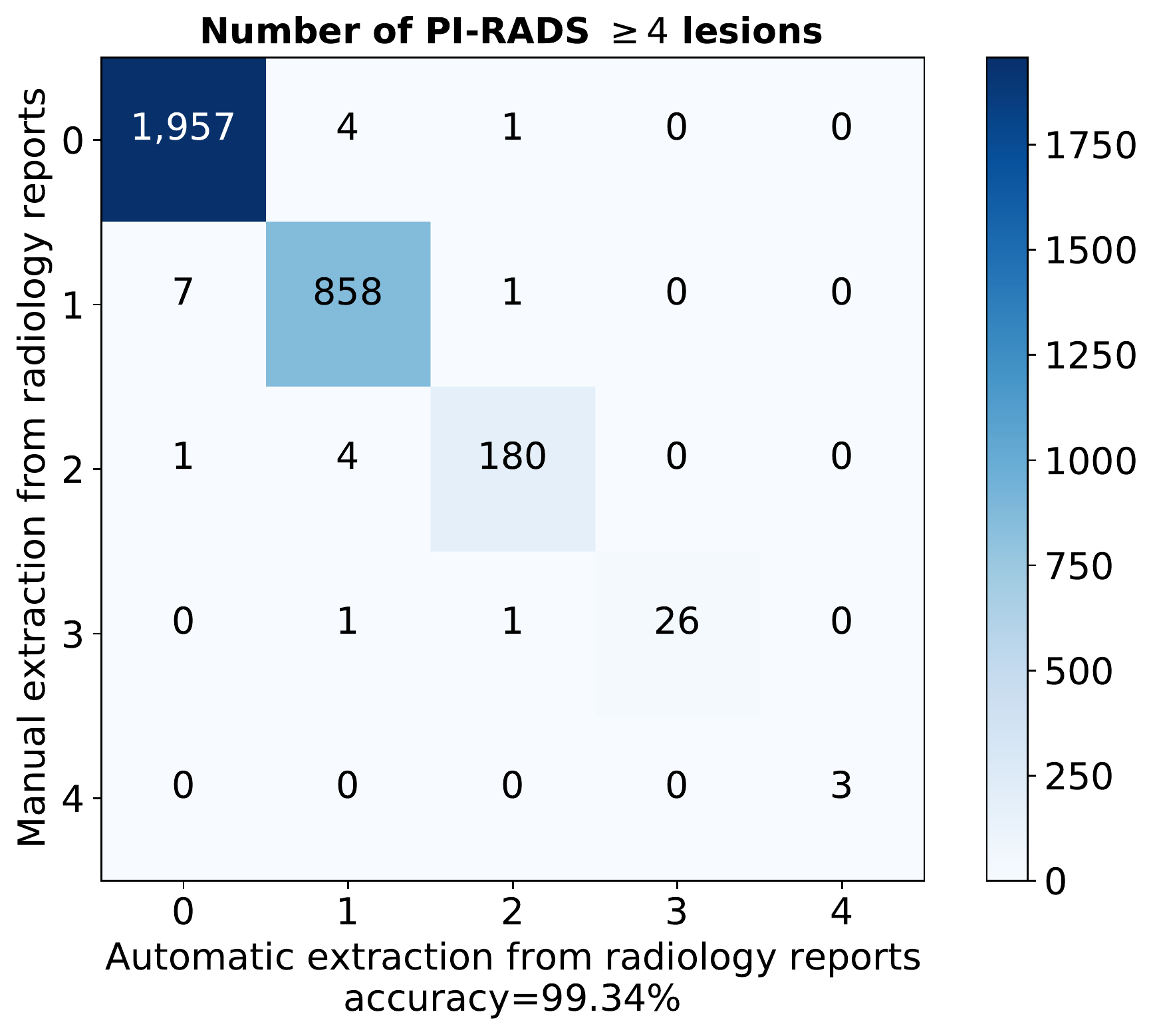}
    \caption{Confusion matrix for number of clinically significant findings in a radiology report. Evaluated on the labelled \nameRUMC\ subset. 
    }
    \label{fig:automatic-finding-extraction-performance}
\end{figure}

Our score extraction script correctly identified the number of clinically significant lesions for \perfAutomaticFindingExtractionTotalCorrect\ out of the \perfAutomaticFindingExtractionTotal\ (\perfAutomaticFindingExtractionAccuracy) radiology reports in our manually labelled \nameRUMC\ dataset. 
We excluded reports and their studies when no PI-RADS scores could be extracted from the report:  \perfAutomaticFindingExtractionTotalExcludedLabelled\ cases (\perfAutomaticFindingExtractionTotalExcludedPercentage) from the labelled \nameRUMC\ dataset and \perfAutomaticFindingExtractionTotalExcludedUnlabelled\ cases (\perfAutomaticFindingExtractionTotalExcludedUnlabelledPercentage) from the unlabelled \nameRUMC\ dataset. 
Full breakdown of automatically extracted versus manually determined number of significant lesions is given in \Cref{fig:automatic-finding-extraction-performance}. 
Typing mistakes and changed scores in the addendum were the main source of the \perfAutomaticFindingExtractionTotalIncorrect\ (\perfAutomaticFindingExtractionErrorRate) incorrect extractions, 
which is an error rate similar to what we observed for our annotators. 


\subsection{Localisation of Report Findings}
\begin{figure}[h!]
    \centering
    \includegraphics[width=\columnwidth]{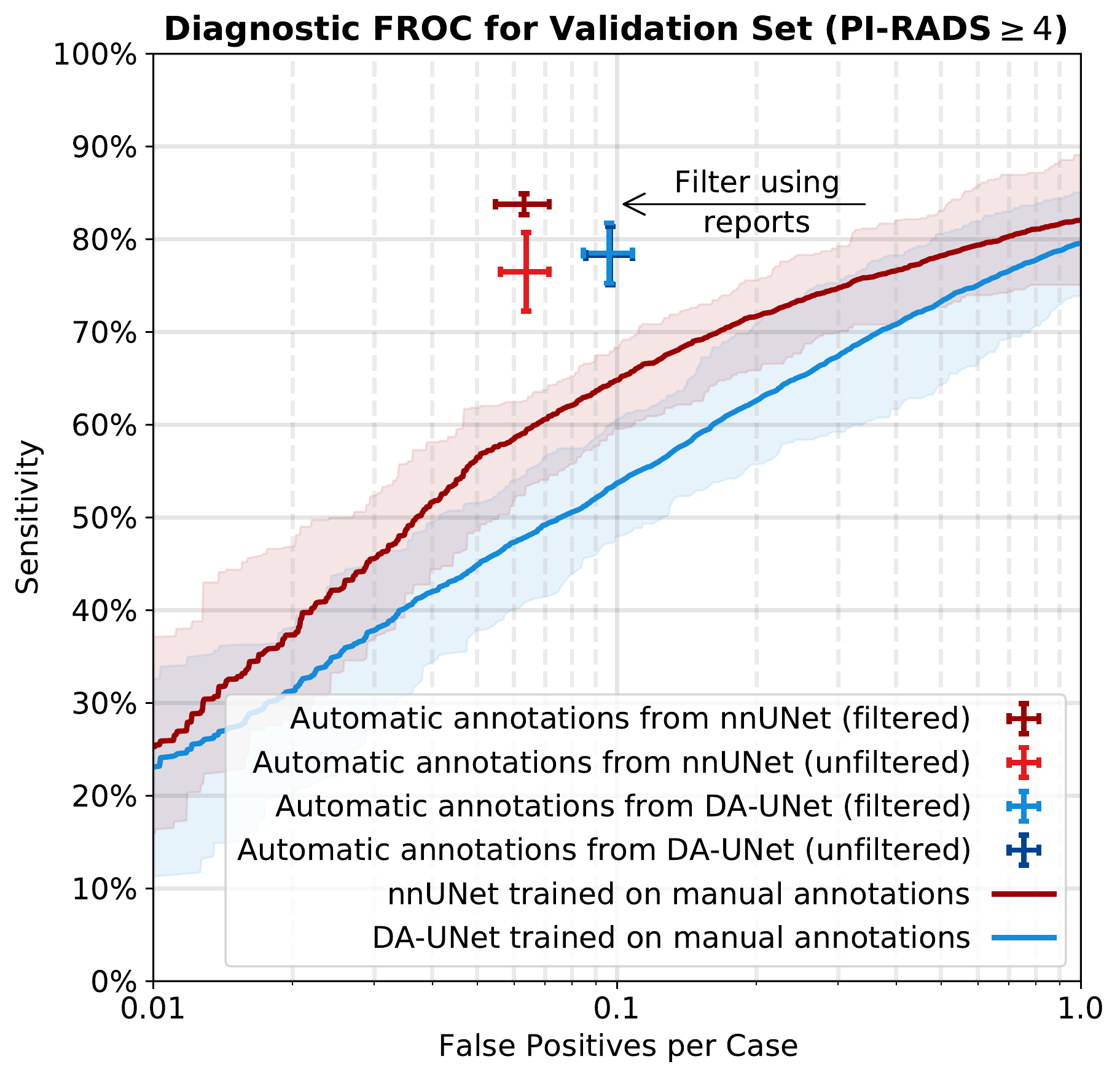}
    \caption{Free-Response Receiver Operating Characteristic (FROC) curves for matching manually annotated PI-RADS $\geq 4$ lesions. Shaded areas indicate the $95\%$ confidence intervals. Masking model predictions with radiology reports significantly improves the false positive rate at high sensitivity, which is the useful range for annotations. }
    \label{fig:AVA_FROC}
\end{figure}

Both prostate cancer segmentation architectures, \namennUNet\ and \nameUNetagpp, can achieve high detection sensitivity. 
At this high sensitivity operating point, the models also propose a large number of false positive lesion candidates, as indicated by the Free-Response Receiver Operating Characteristic (FROC) curve shown in \Cref{fig:AVA_FROC}. 
Masking the models' lesion candidates 
with the number of clinically significant report findings, \nsig, greatly reduces the number of false positive lesion candidates. 
At the sensitivity of the unfiltered automatic annotations, masking with radiology reports reduced the average number of false positives per case from \perfLesionLocalisationnnUNetFPsModelUnfiltered\ to \perfLesionLocalisationnnUNetFPsUnfiltered\ for \namennUNet\ and from \perfLesionLocalisationUNetagppFPsModelUnfiltered\ to 
\perfLesionLocalisationUNetagppFPsUnfiltered\ for \nameUNetagpp\ \added{(trained with 5-fold cross-validation on \numManualDelineationsFullDataset\ manually annotated exams)}. This more than five-fold reduction in false positives greatly increases the usability of the automatic annotations. 


Studies where we could extract fewer than \nsig\ lesion candidates were excluded. 
This excludes studies where we are certain to miss lesions, and thus increases sensitivity. 
From the automatic annotations from \namennUNet\ we excluded \numLesionsExcludednnUNet\ studies, resulting in a sensitivity of  \perfLesionLocalisationnnUNetSensitivityFiltered\ at \perfLesionLocalisationnnUNetFPsFiltered\ false positives per study. 
From the automatic annotations from \nameUNetagpp\ we excluded \numLesionsExcludedUNetagpp\ studies, resulting in a sensitivity of 
\perfLesionLocalisationUNetagppSensitivityFiltered\ at \perfLesionLocalisationUNetagppFPsFiltered\ false positives per study.

\subsection{Segmentation of Report Findings}
Spatial similarity between the automatic and manual annotations is good. 
\added{Trained with 5-fold cross-validation on \numManualDelineationsFullDataset\ manual annotations, }
\namennUNet\ achieved a Dice similarity coefficient (DSC) of \perfLesionSegmentationAVAnnUNetMatched, and \nameUNetagpp\ achieved \perfLesionSegmentationAVAUNetagppMatched\ DSC. Including the missed manual annotations as a DSC of zero, reduces this to \perfLesionSegmentationAVAnnUNet\ for \namennUNet\ and \perfLesionSegmentationAVAUNetagpp\ for \nameUNetagpp. 
The full distribution of DSC against lesion volume is given in the Supplementary Materials. 

\Cref{fig:pseudo-label-pipeline} shows automatic annotations from \namennUNet, with a DSC of \DSCPatientOneTop\ ($\approx$ mean) for the upper lesion of Patient 1, a DSC of \DSCPatientTwo\ ($\approx$ mean $+$ std.) for Patient 2 and a DSC of \DSCPatientThree\ ($\approx$ mean $-$ std.) for Patient 3. 

\subsection{Prostate Cancer Detection}

\deleted{\subsubsection{Automatic Labels Compared to Manual Labels}}
\deleted{Training on automatic labels resulted in models with equal or better performance, compared to training on the same number of manual labels. For the external test set with histopathology-confirmed ground truth, \namennUNet\ achieved a case-level AUROC of \perfZGTAUCnnUNetAutomaticSubsetOnly\ when training on automatically annotated studies and \perfZGTAUCnnUNetManual\ when training on manually annotated studies (\statZGTAUCnnUNetManualBetterThanAutomaticOnly). Training \nameUNetagpp\ on automatically annotated studies resulted in a case-level AUROC of \perfZGTAUCUNetagppAutomaticSubsetOnly, significantly outperforming training on manually annotated exams, which achieved \perfZGTAUCUNetagppManual\ AUROC (\statZGTAUCUNetagppManualBetterThanAutomaticOnly). }

\deleted{Lesion detection performance, as measured by the \namepAUC, was \perfZGTpAUCnnUNetManual\ for \namennUNet\ when training on automatically annotated studies and \perfZGTpAUCnnUNetAutomaticSubsetOnly\ when training on manually annotated studies  (\statZGTpAUCnnUNetManualBetterThanAutomaticOnly). \nameUNetagpp\ trained on automatically annotated exams achieved \perfZGTpAUCUNetagppAutomaticSubsetOnly\ \namepAUC, significantly outperforming training on manually annotated exams, which achieved \perfZGTpAUCUNetagppManual\ \namepAUC\ (\statZGTpAUCUNetagppManualBetterThanAutomaticOnly). 
Sensitivity at one false positive per case increased from \perfZGTSensAtOneFPnnUNetManual\ to \perfZGTSensAtOneFPnnUNetAutomaticSubsetOnly\ (\statZGTSensAtOneFPnnUNetManualBetterThanAutomaticOnly) for \namennUNet\ and from \perfZGTSensAtOneFPUNetagppManual\ to \perfZGTSensAtOneFPUNetagppAutomaticSubsetOnly\ (\statZGTSensAtOneFPUNetagppManualBetterThanAutomaticOnly) for \nameUNetagpp. 
See the top row of Figure 6 for the ROC and FROC curves.}

\subsubsection{\replaced{Training with Report-guided Automatic Annotations}{\\Augmenting Training Set with Automatic Annotations}}
\replaced{Semi-supervised training with \numManualDelineationsFullDataset\ manually annotated and \numUnlabelledStudiesFullDataset\ automatically annotated exams significantly improved model performance, compared to supervised training with \numManualDelineationsFullDataset\ manually annotated exams. }{%
Augmenting the training set with automatically annotated exams significantly improved model performance.}
For the external test set with histopathology-confirmed ground truth, the case-\replaced{based}{level} AUROC increased from \perfZGTAUCnnUNetManual\ to \perfZGTAUCnnUNetManualAndAutomatic\ (\statZGTAUCnnUNetManualBetterThanManualAndAutomatic) for \namennUNet\ and from \perfZGTAUCUNetagppManual\ to \perfZGTAUCUNetagppManualAndAutomatic\ (\statZGTAUCUNetagppManualBetterThanManualAndAutomatic) for \nameUNetagpp. 

On this external test set, \added{a consensus of}\ experienced radiologists had a sensitivity of \perfZGTSensitivityRadiologist\ at \perfZGTSpecificityRadiologist\ specificity. 
At the same sensitivity, adding \replaced{automatically annotated exams}{automatic annotations} improved the model's specificity from \perfZGTSpecificitynnUNetManual\ to \perfZGTSpecificitynnUNetManualAndAutomatic\ (\statZGTSpecificitynnUNetManualBetterThanManualAndAutomatic) for \namennUNet\ and from \perfZGTSpecificityUnetagppManual\ to \perfZGTSpecificityUnetagppManualAndAutomatic\ (\statZGTSpecificityUNetagppManualBetterThanManualAndAutomatic) for \nameUNetagpp. 

\begin{figure*}[h!]
    \centering
    \includegraphics[width=\linewidth]{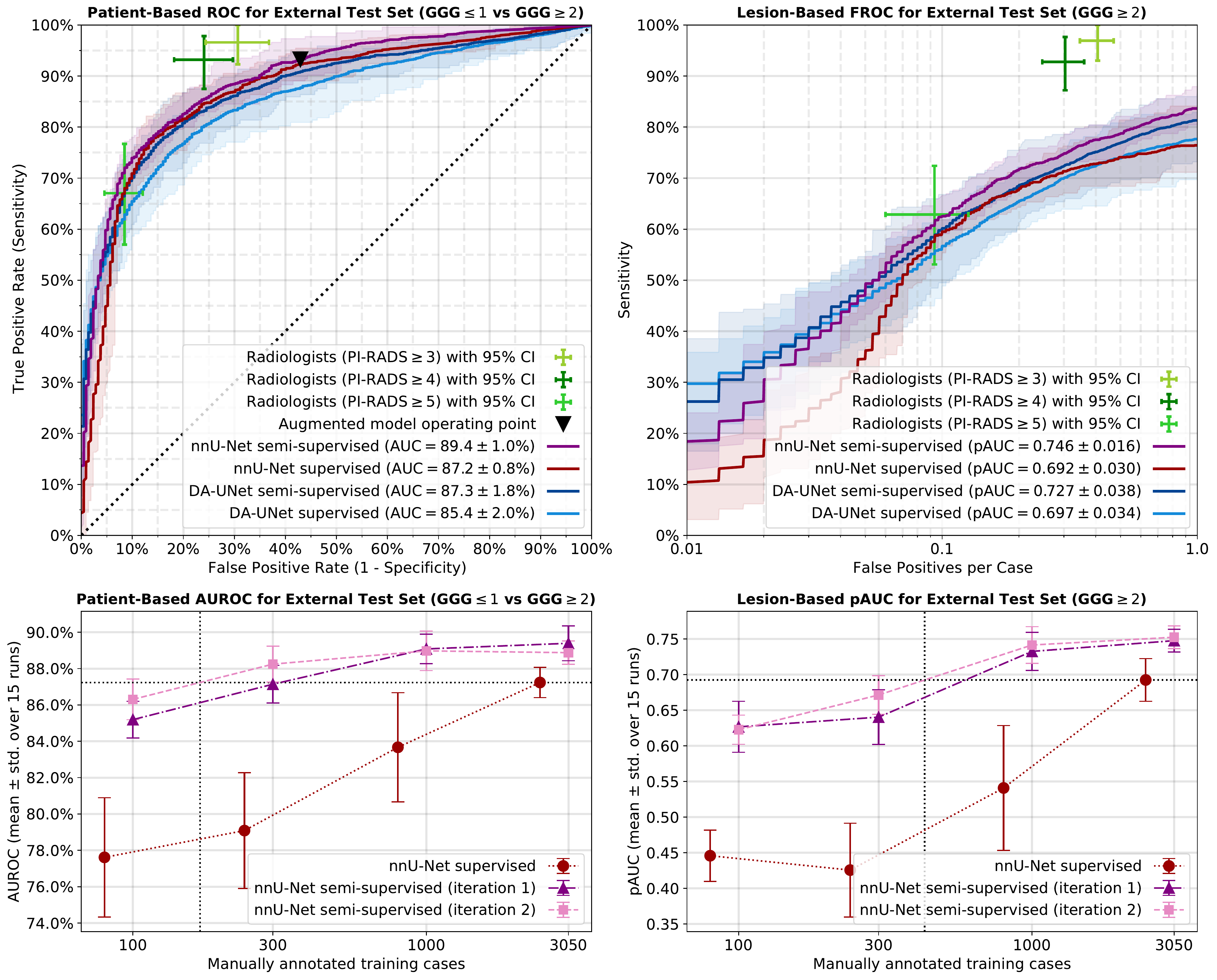}
    
    \caption{
    \replaced{\panelstyle{(top row)}\ Model performance for (semi-)supervised training. Supervised models are trained with 5-fold cross-validation on \numManualDelineationsFullDataset\ manually annotated exams, and self-training also includes automatically annotated exams from \numUnlabelledStudiesFullDataset\ unlabelled exams. Automatic annotations for self-training are generated using the supervised models with the same architecture. 
    \panelstyle{(bottom row)}\ 
    \namennUNet\ performance for \numManualDelineationsFirstSubset, \numManualDelineationsSecondSubset, \numManualDelineationsThirdSubset\ or \numManualDelineationsFullDataset\ manually annotated exams, combined with \numUnlabelledStudiesFirstSubset, \numUnlabelledStudiesSecondSubset, \numUnlabelledStudiesThirdSubset\ or \numUnlabelledStudiesFullDataset\ unlabelled cases, respectively. Automatic annotations for iteration 1 are generated using the supervised models, automatic annotations for iteration 2 are generated using the semi-supervised models from iteration 1. }{
    Model performance for 
    \panelstyle{(top row)}\ training on manual PI-RADS $\geq 4$ lesion annotations versus the same number of automatic annotations and 
    \panelstyle{(bottom row)}\ training on manual annotations versus manual+automatic annotations generated using \namennUNet\ or \nameUNetagpp. }
    The \panelstyle{(left)}\ panels show ROC \replaced{performance}{curves} for patient-based diagnosis of exams with at least one Gleason grade group (GGG) $\geq 2$ lesion, and the \panelstyle{(right)}\ panels show FROC \replaced{performance}{curves} for lesion-based diagnosis of GGG$ \geq 2$ lesions. 
    All models are trained on radiology based PI-RADS $\geq 4$ annotations and evaluated on the external test set with histopathology-confirmed ground truth.
    \color{black}Shaded areas indicate the $95\%$ confidence intervals \replaced{from}{over} \added{15 or} 25 \replaced{independent training runs}{restarts}. 
    }
    \label{fig:ZGT_four}
\end{figure*}

Detection and risk stratification of individual lesions also benefitted from including the automatically labelled training samples. 
Model performance, as measured by the \namepAUC, increased from \perfZGTpAUCnnUNetManual\ to \perfZGTpAUCnnUNetManualAndAutomatic\ (\statZGTpAUCnnUNetManualBetterThanManualAndAutomatic) for \namennUNet\ and from \perfZGTpAUCUNetagppManual\ to \perfZGTpAUCUNetagppManualAndAutomatic\ (\statZGTpAUCUNetagppManualBetterThanManualAndAutomatic) for \nameUNetagpp. 
Sensitivity at one false positive per case increased from \perfZGTSensAtOneFPnnUNetManual\ to \perfZGTSensAtOneFPnnUNetManualAndAutomatic\ (\statZGTSensAtOneFPnnUNetManualBetterThanManualAndAutomatic) for \namennUNet\ and from \perfZGTSensAtOneFPUNetagppManual\ to \perfZGTSensAtOneFPUNetagppManualAndAutomatic\ (\statZGTSensAtOneFPUNetagppManualBetterThanManualAndAutomatic) for \nameUNetagpp. 
See the \replaced{top}{bottom} row of \Cref{fig:ZGT_four} for the ROC and FROC curves.

\newcommand\panelwidth{.94\columnwidth}

\subsection{\added{Annotation-efficiency of Training with Report-guided Automatic Annotations}}
\added{Semi-supervised training significantly increased model performance for all investigated manual annotation budgets, compared to supervised training with the same number of manually annotated exams (\maxPvalueIterationOneBetterThanSupervised\ for each budget). Iteration 2 of semi-supervised training, where automatic annotations were generated by ensembling models from iteration 1 of semi-supervised training, generally performed better than iteration 1 of semi-supervised training. However, only AUROC improvement for manual annotation budgets of \numManualDelineationsFirstSubset\ and \numManualDelineationsSecondSubset\ exams were statistically significant (\PvalueIterationTwoBetterThanIterationOneFirstSubsetAUROC\ and \PvalueIterationTwoBetterThanIterationOneSecondSubsetAUROC). ROC and FROC performances, summarised to the AUROC and pAUC, are shown in the bottom row of \Cref{fig:ZGT_four}.


Semi-supervised training (iteration 2) with \numManualDelineationsSecondSubset\ manual annotations exceeded case-based AUROC performance of supervised training with \numManualDelineationsFullDatasetPerSupervisedRun\ manually annotated exams (\PvalueSupervisedFullBetterThanThreeHundredIterationTwoAUROC). Performance of semi-supervised training with \numManualDelineationsFirstSubset\ manually annotated exams came close (\PvalueOneHundredIterationTwoBetterThanSupervisedFullAUROC). Interpolation suggests the supervised performance is matched with \numManualDelineationsMatchedPerformanceFullDatasetAUROC\ manually annotated exams (\annotationEfficiencyMatchedPerformanceFullDatasetAUROC\ more annotation-efficient). 

Semi-supervised training (iteration 2) with \numManualDelineationsThirdSubset\ manually annotated exams exceeded lesion-based pAUC performance of supervised training with \numManualDelineationsFullDatasetPerSupervisedRun\ manual annotations (\PvalueSupervisedFullBetterThanOneThousandIterationTwopAUC). Performance of semi-supervised training with \numManualDelineationsSecondSubset\ manually annotated exams came close (\PvalueThreeHundredIterationTwoBetterThanSupervisedFullpAUC). Interpolation suggests the supervised performance is matched with \numManualDelineationsMatchedPerformanceFullDatasetpAUC\ manually annotated exams (\annotationEfficiencyMatchedPerformanceFullDatasetpAUC\ more annotation-efficient). 
}
\section{Discussion and Conclusion}\color{black}
\label{sec:discussion}
\replaced{Semi-supervised training significantly improved patient-based risk stratification and lesion-based detection performance of our prostate cancer detection models for all investigated manual annotation budgets, compared to supervised training with the same number of manually annotated exams.}{Patient-level risk stratification and lesion-level detection and risk stratification of our prostate cancer detection models significantly improved by augmenting the training set with automatically-generated prostate cancer annotations guided by radiology report findings.} 
This improved performance demonstrates the feasibility of our \replaced{semi-supervised training method with report-guided automatic annotations.}{annotation method.} 
Furthermore, the automatic annotations are of sufficient quality to speed up the manual annotation process, by identifying negative cases that do not need to be looked at, and by providing high quality segmentation masks for the majority of the positive cases. 
%
Our \replaced{semi-supervised training method}{automatic labelling procedure} enabled us to utilise thousands of additional prostate MRI exams with radiology reports from clinical routine, without manually annotating each finding in the MRI scan. 
\deleted{While automatic annotations reflect the radiologist's findings less accurately compared to human annotators, we found that training exclusively on automatically annotated exams did not lead to statistically inferior model performance, compared to training on an equal number of manual annotations. For \nameUNetagpp\ performance even increased, for both case-level risk stratification and lesion-level detection.}
\deleted{We believe the 
non-inferior or improved performance could be caused by a higher level of consistency in the automatic annotations, compared to the manual annotations. Manual annotations reflect the high inter- and intra-reader variability of PI-RADS $\geq 4$ findings \citep{rosenkrantz2016interobserver, smith2019intra}, while the automatic annotations are generated using a single ensemble of models. 
Increased consistency could result in a more stable training signal, helping model convergence. 
The larger improvement of \nameUNetagpp\ compared to \namennUNet\ also raises the question which components caused this difference. Automatic annotations generated using a model trained with Focal Loss ($\alpha = 0.75$) could be better than automatic annotations generated using a model trained with cross-entropy. Or, a model trained with Focal Loss ($\alpha = 0.75$) may benefit more from the increased consistency of automatic annotations. Or, another difference between the training setup of \namennUNet\ and \nameUNetagpp\ causes this difference in performance increase. Further research is necessary to answer these questions, and push performance with automatically annotated exams further. }

\replaced{Semi-supervised training}{Augmenting the training dataset} with automatically labelled prostate MRI exams consistently improved model performance, with a group of baseline and augmented models truly reflecting the difference in performance due to the automatically labelled exams. Comparing groups of models \deleted{with a permutation test} ensures results are not due to variation in performance inherent to deep learning's stochastic nature\footnote{Sources of variation include the model's random initialisation, order of training batches and data augmentations, resulting in differences in model performance between training runs. }. 


Automatically annotating training samples enables us to improve our models, but also speeds up the manual labelling process for new cases. Accurate PI-RADS score extraction from radiology reports enables us to automatically identify all negative cases (with $\leq 1\%$ error rate), saving the repetitive operation of reading the radiology report only to designate a study as negative. As this entails approximately $60\%$ of the studies
, this already amounts to a large time saving. 
Furthermore, the segmentation masks from \namennUNet\ are often of sufficient quality 
to only require verification of the location, saving significant amounts of time for positive studies as well. 

Direct applicability of \deleted{both} the automatic PI-RADS extraction from radiology reports \deleted{and the models for the automatic annotation procedure} \replaced{is}{are}, however, limited. The rule-based score extraction was developed with the report templates from \nameRUMC\ in mind, and is likely to fail for reports with a different structure. For institutes that also have structured reports, the rule-based score extraction can be adapted to match their findings. For unstructured (free text) reports, the task of counting the number of clinically significant findings can be performed by a deep learning-based natural language processing model. This model can be trained on the reports of the manually labelled subset, using the number of findings found in the manual annotations as labels. 

Another limitation is that the prostate cancer detection models were trained with prostate MRI scans from a single vendor (Siemens Healthineers, Erlangen; Magnetom Trio/Skyra/Prisma/Avanto). Therefore, these models are likely to perform inferior on scans from different scanner models. 

In conclusion, \replaced{semi-supervised training with report-guided automatic annotations significantly improved csPCa detection performance}{the report-guided automatic annotations are of high quality}, allowing unlabelled samples to be leveraged without additional manual effort. 
Furthermore, automatic annotations can speed up the manual annotation process. 
Our proposed method is widely applicable, paving the way towards larger datasets with equal or reduced annotation time.

\bibliographystyle{model2-names.bst}
\biboptions{authoryear}
\bibliography{refs}

\pagenumbering{roman}
\renewcommand{\theHfigure}{Sup. Mat.\thetable}
\renewcommand\thefigure{\Alph{figure}}
\renewcommand\thesection{\Alph{section}}
\renewcommand\thetable{\Alph{table}}
\setcounter{section}{0}
\setcounter{figure}{0} 

\section*{Supplementary Materials}

\section{Study Population}
\label{sec:study_population}
\setlength\heavyrulewidth{0.5ex}
\setlength\lightrulewidth{0.3ex}
\renewcommand\midrule{\specialrule{\lightrulewidth}{0pt}{0pt}}
\renewcommand\toprule{\specialrule{\heavyrulewidth}{0pt}{0pt}}
\renewcommand\bottomrule{\specialrule{\heavyrulewidth}{0pt}{0pt}}
\renewcommand{\arraystretch}{1.4} 
\definecolor{lightgray}{gray}{0.9}
\newcommand\secFootnote{^{\text{§}}}
\newcommand\stardwidth{0.4\textwidth}
\newcommand{\rowcol}{\rowcolor[gray]{0.93}[\tabcolsep][\tabcolsep]}
\begin{figure}[H]
    \centering
    \begin{tabular}{cc}
\begin{minipage}{0.54\textwidth}
\begin{threeparttable}
\begin{tablenotes}
\small
\item \panelstyle{Table A.} ~~ Patient demographic and characteristics for datasets acquired at Radboud University Medical Centre (\nameRUMC) and Ziekenhuisgroep Twente (ZGT). Characteristic values are followed by their interquartile range (IQR), if applicable. 
\vspace{0.05em} 
\end{tablenotes}
\begin{tabular}{lcc} \toprule
    {\textbf{Characteristic}}                & {\textbf{\nameRUMC}}           &  {\textbf{ZGT}}         \\ \midrule
    Number of Patients                          & $\numbpMRIPatientsFullDataset$ &  $\numZGTPatients$      \\ \midrule
    Number of Studies                           & $\numbpMRIVisitsFullDataset$   &  $\numZGTVisits$        \\ \midrule
    \rowcol • Benign                         & $4,734\secFootnote$            &  $212$                  \\ \hline
    \rowcol • Malignant $(\geq 1 \text{ csPCa}^{*})$  &  $3,022\secFootnote$  &   $88$                  \\ \midrule
    Median PSA (ng/mL)                       & \distPSALevelsRUMCShort        &  \distPSALevelsZGTShort \\ \midrule
    Median Age (years)                       & \distAgeRUMCShort              &  \distAgeZGTShort       \\ \midrule
    Median Prostate Volume (cm$^3$)          & $64\ (46-91)$                  &  $50\ (40-69)$          \\ \midrule
    \multicolumn{2}{l}{MRI Scanners ($3$ T, Surface Coils)}                   & ~                       \\ \midrule
    \rowcol • Magnetom Trio/Skyra$^+$        & $88.9\%$                       & $100\%$                 \\ \hline
    \rowcol • Magnetom Prisma$^+$            & $11.0\%$                       & $-$                     \\ \hline
    \rowcol • Magnetom Avanto$^+$ $(1.5\ $T$)$ & $0.1\%$                      & $-$                     \\ \midrule
    T2W Acquisition                          & ~                              & ~                       \\ \midrule
    \rowcol • In-Plane Resolution (mm/voxel) & $0.30 \pm 0.08$                & $0.50 \pm 0.00$         \\ \hline
    \rowcol • Slice Thickness (mm/voxel)     & $3.60 \pm 0.20$                & $3.00 \pm 0.00$         \\ \midrule
    DWI/ADC Acquisition                      & ~                              & ~                       \\ \midrule
    \rowcol • In-Plane Resolution (mm/voxel) & $2.00 \pm 0.05$                & $2.00 \pm 0.00$         \\ \hline
    \rowcol • Slice Thickness (mm/voxel)     & $3.60 \pm 0.20$                & $3.00 \pm 0.00$         \\ \hline
    \rowcol • Computed High b-Value          & b1400                          & b1400                   \\ \midrule
    MRI-Detected Lesions                     & $10,564\secFootnote$           & $464$                   \\ \midrule
    \rowcol • PI-RADS $\leq 2$               & $5,958\secFootnote$            & $248$                   \\ \hline
    \rowcol • PI-RADS $3$                    & $983\secFootnote$              & $35$                    \\ \hline
    \rowcol • PI-RADS $4$                    & $2,115\secFootnote$            & $92$                    \\ \hline
    \rowcol • PI-RADS $5$                    & $1,508\secFootnote$            & $89$                    \\ \midrule
    \multicolumn{2}{l}{Histopathology-Confirmed Lesions \hspace{19pt} N/A}    & $191$                   \\ \midrule
    \rowcol • GGG $1$ (GS $\leq 3+3$)        & N/A                            & $94$                    \\ \hline
    \rowcol • GGG $2$ (GS $3+4$)             & N/A                            & $63$                    \\ \hline
    \rowcol • GGG $3$ (GS $4+3$)             & N/A                            & $11$                    \\ \hline
    \rowcol • GGG $4$ (GS $4+4$)             & N/A                            & $5$                     \\ \hline
    \rowcol • GGG $5$ (GS $\geq 4+5$)        & N/A                            & $18$                    \\ \bottomrule
\end{tabular}
\begin{tablenotes}
  \small
  \item $^*$ \nameRUMC\ csPCa: PI-RADS $\geq 4$; ZGT csPCa: GGG $\geq 2$
  \item $^+$ Siemens Healthineers, Erlangen, Germany
  \item $\secFootnote$ Determined semi-automatically
\end{tablenotes}
\end{threeparttable}
\end{minipage} & \begin{minipage}{0.42\textwidth} 
    \centering
    \begin{tabular}{cc}
         \includegraphics[width=\textwidth]{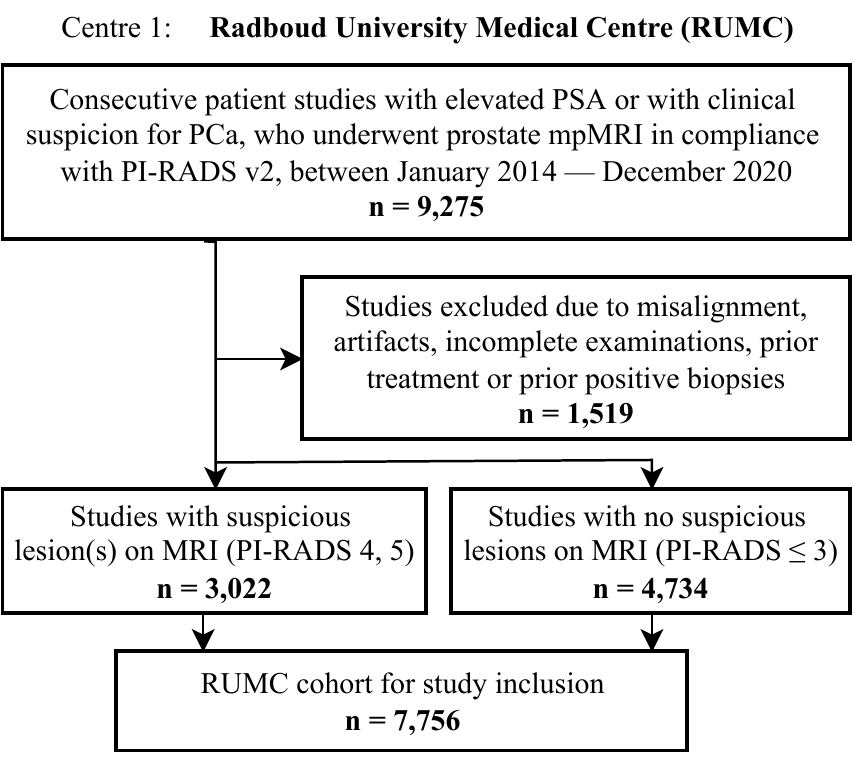} \\
         ~ \\
         \includegraphics[width=\textwidth]{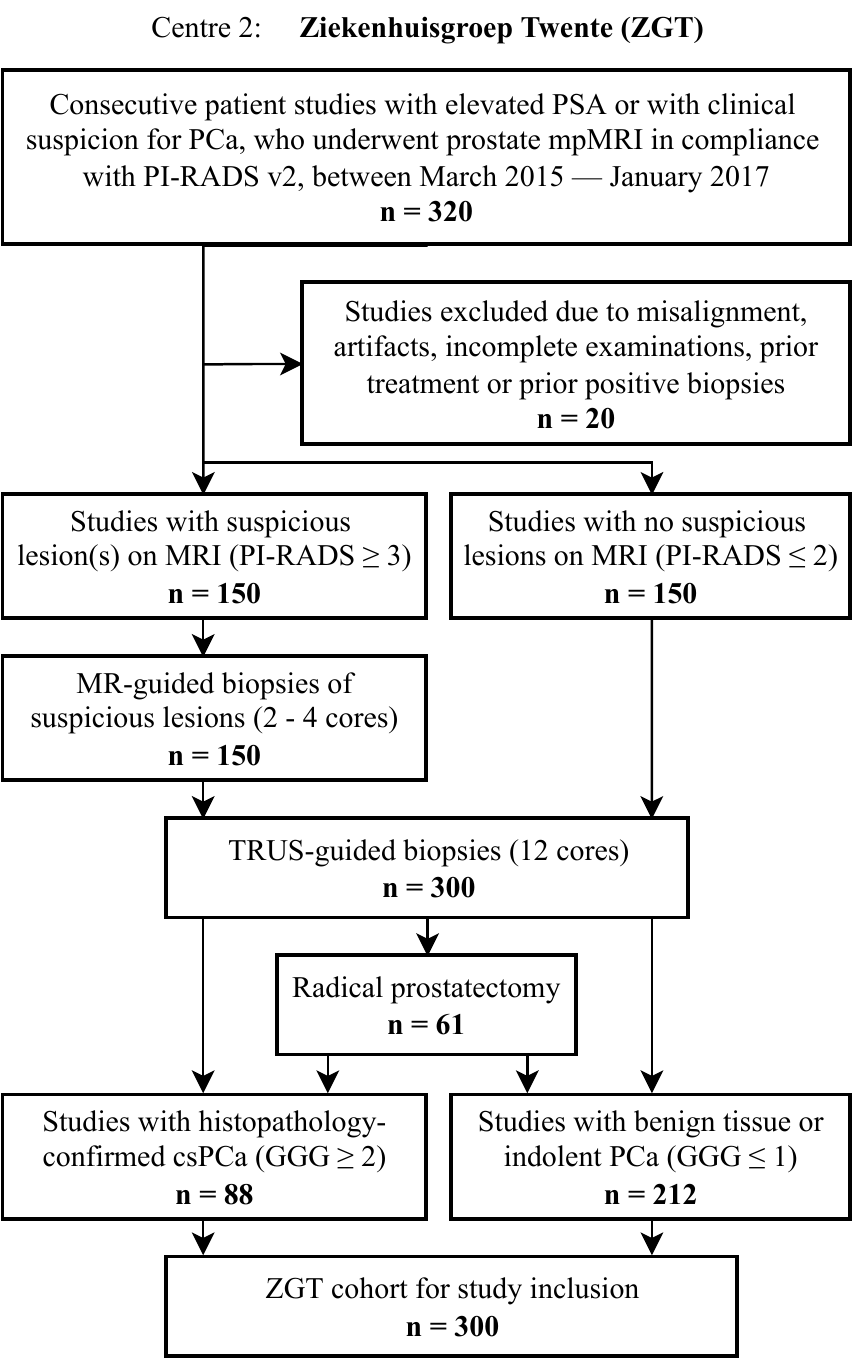}
    \end{tabular}
    
    \caption{STARD diagram for study inclusion/exclusion criteria applied across both centres: Radboud University Medical Centre (\nameRUMC) and Ziekenhuisgroep Twente (ZGT).}
\end{minipage}

    \end{tabular}
\end{figure}
\begin{figure*}[h!]
\begin{minipage}[t]{0.48\textwidth}
    \input{supplementary-materials/2a-parse-report}
\end{minipage}%
\hfill
\begin{minipage}[t]{0.48\textwidth}
    \input{supplementary-materials/2b-detection-architecture}
\end{minipage}%
\end{figure*}

\begin{figure*}[h!]
  \centering\includegraphics[width=\linewidth]{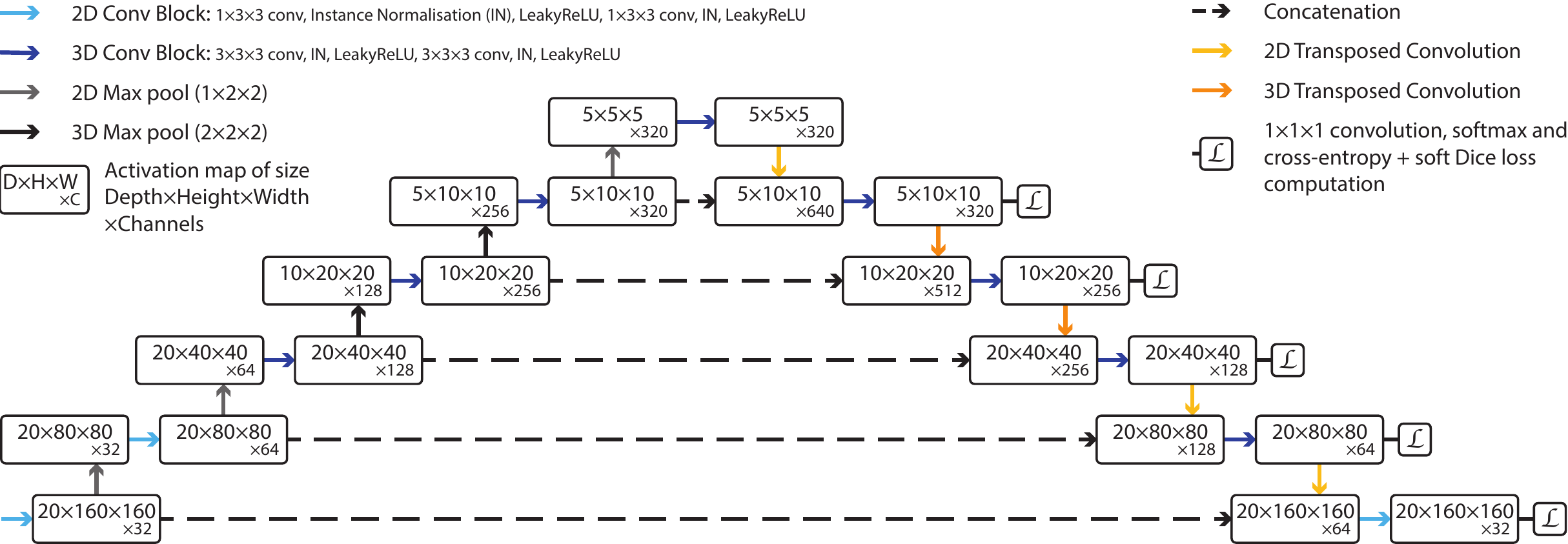}\par
  \caption{Architecture schematic of the \namennUNet, as configured for our prostate cancer segmentation task. }
    \label{fig:detector_networks}
\end{figure*}%

~\newpage~\newpage~\newpage~\newpage
\FloatBarrier
\section{Lesion Segmentation Quality}
\begin{figure}[h!]
    \centering
    \vspace{0pt}
    \includegraphics[width=\columnwidth]{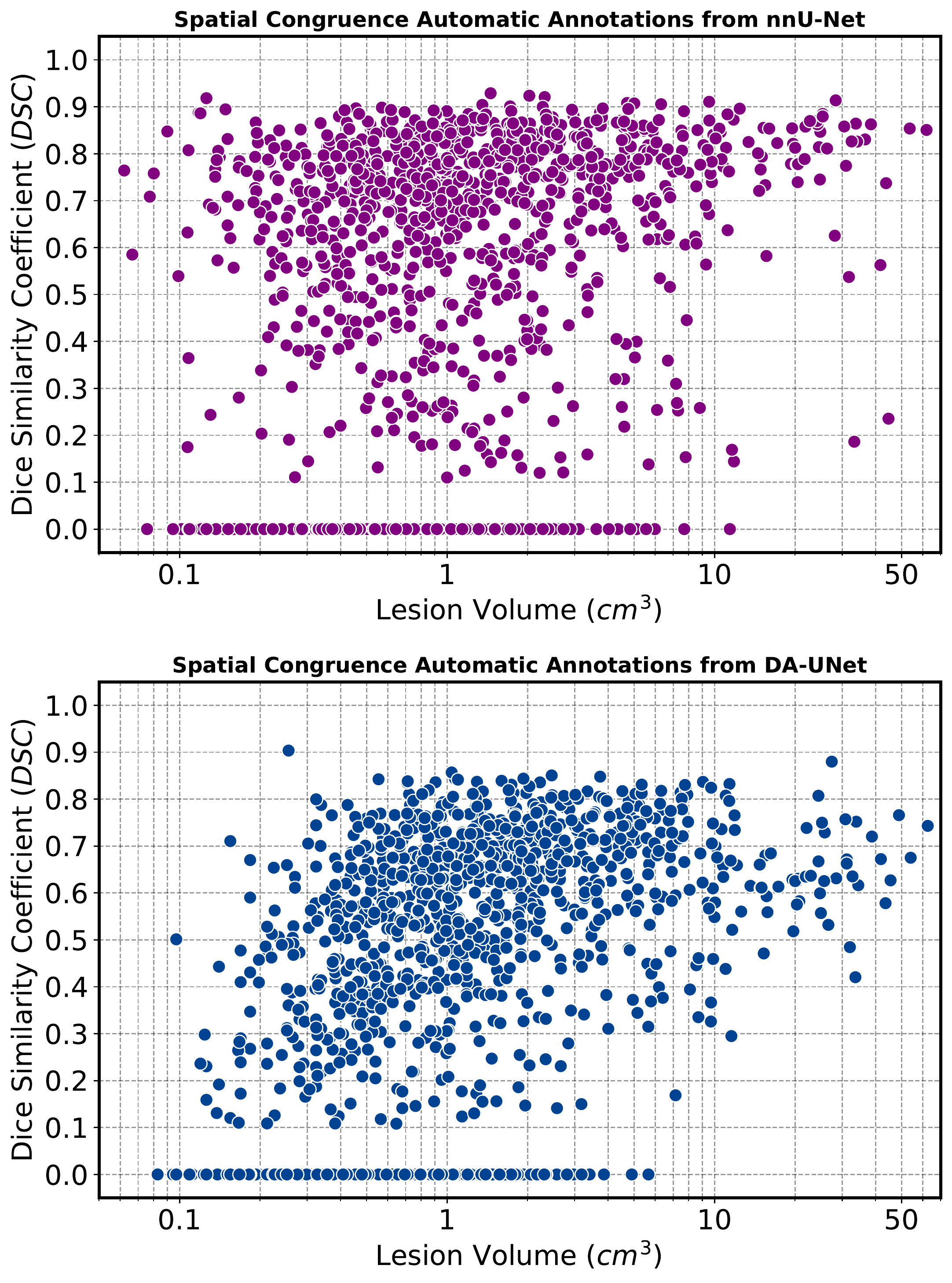}
    \caption{Spatial congruence between automatic and manual csPCa annotations, as measured by the Dice similarity coefficient, for automatic annotations derived from \panelstyle{(top)}\ \namennUNet\ or \panelstyle{(bottom)}\ \nameUNetagpp. 
    Both methods were evaluated on the labelled \nameRUMC\ dataset with 5-fold cross-validation, excluded studies due to empty PI-RADS extraction from the radiology report, and excluded studies with insufficient lesion candidates. All metrics were computed in 3D. }
    \label{fig:AVA-quality-DSC}
\end{figure}

\end{document}